\documentclass[useAMS,usenatbib]{mnras}
\usepackage{graphics,epsfig,psfig}
\usepackage[normalem]{ulem}
\usepackage{xcolor}
\usepackage{commath}
\usepackage{float}
\usepackage{subfig}
\usepackage{graphicx}
\usepackage{amsmath, bm}
\usepackage[]{inputenc,amssymb}
\def \be{\begin{equation}}
\def \ee{\end{equation}}
\def \bea{\begin{eqnarray}}
\def \eea{\end{eqnarray}}

\definecolor{webgreen}{rgb}{0,.5,0}
\definecolor{webbrown}{rgb}{.6,0,0}

\definecolor{cadmiumgreen}{rgb}{0.0, 0.42, 0.24}

\title[HI IM with MIGHTEE]{HI intensity mapping with the MIGHTEE survey: power spectrum estimates}
\author[]{Sourabh Paul${^1}$ \thanks{sourabh.paul@gmail.com}, Mario G. Santos$^{1,2}$, Junaid Townsend${^1}$, Matt J. Jarvis${^{3,1}}$,
\newauthor
Natasha Maddox$^4$, Jordan D. Collier${^{5,6}}$, Bradley S. Frank${^{2,5,7}}$, Russ Taylor$^8$\\
$^{1}$ Department of Physics and Astronomy, University of the Western Cape, Robert Sobukhwe Road, Bellville, 7535, South Africa\\
$^{2}$ South African Radio Observatory (SARAO), 2 Fir Street, Observatory, Cape Town, 7925, South Africa\\
$^{3}$ Astrophysics, Department of Physics, University of Oxford, Keble Road, Oxford, OX1 3RH, UK\\
$^{4}$ Faculty of Physics, Ludwig-Maximilians-Universit{\"a}t, Scheinerstr. 1, 81679 Munich, Germany\\
$^{5}$ Department of Astronomy, University of Cape Town, Private Bag X3, Rondebosch 7701, South Africa\\
$^{6}$ School of Science, Western Sydney University, Locked Bag 1797, Penrith, NSW 2751, Australia \\
$^{7}$ The Inter-University Institute for Data Intensive Astronomy (IDIA), and \\\hspace{10pt}University of Cape Town, Private Bag X3, Rondebosch, 7701, South Africa \\
$^{8}$ Inter-University Institute for Data Intensive Astronomy, and \\\hspace{10pt}Department of Physics and Astronomy, University of the Western Cape, Robert Sobukhwe Road, Bellville, 7535, South Africa \\
}

\begin{document}

\maketitle

\label{firstpage}

\begin{abstract}
Intensity mapping (IM) with neutral hydrogen is a promising avenue to probe the large scale structure of the Universe. In this paper, we demonstrate that using the 64-dish MeerKAT radio telescope as a connected interferometer, it is possible to make a statistical detection of HI in the post-reionization Universe. With the MIGHTEE (MeerKAT International GHz Tiered Extragalactic Exploration) survey project observing in the L-band ($856 < \nu < 1712$ MHz, $z < 0.66$), we can achieve the required sensitivity to measure the HI IM power spectrum on quasi-linear scales, which will provide an important complementarity to the single-dish IM MeerKAT observations.
We present a purpose-built simulation pipeline that emulates the MIGHTEE observations and forecast the constraints that can be achieved on the HI power spectrum at $z = 0.27$ for $k > 0.3$ $\rm{Mpc}^{-1}$ using the foreground avoidance method. We present the power spectrum estimates with the current simulation on the COSMOS field that includes contributions from HI, noise and point source models constructed from the observed MIGHTEE data. The results from our \textit{visibility} based pipeline are in qualitative agreement to the already available MIGHTEE data. This paper demonstrates that MeerKAT can achieve very high sensitivity to detect HI with the full MIGHTEE survey on quasi-linear scales (signal-to-noise ratio $> 7$ at $k=0.49$ $\rm{Mpc}^{-1}$) which are instrumental in probing cosmological quantities such as the spectral index of fluctuation, constraints on warm dark matter, the quasi-linear redshift space distortions and the measurement of the HI content of the Universe up to $z\sim 0.5$.

\end{abstract}


\section{Introduction}
The spatial distribution of matter in the large scale structure of the Universe imprints intriguing details of many fundamental quantities imperative to our understanding of the Universe.
However, this matter distribution is not directly observable to us and tracers such as galaxies are needed to map the cosmic web. On large scales where perturbations are small, the clustering properties of the tracers follow the fluctuations of the underlying matter field. Large galaxy surveys such as the Sloan Digital Sky Survey \citep[SDSS,][]{York_2000} have mapped large areas of the sky at low-redshift and aided measurements of the cosmological baryon acoustic oscillation signal \citep[BAO,][]{Eisenstein_2005}. In particular, the anisotropic galaxy clustering measurements have put constraints on various cosmological parameters \citep{Reid_2012, Chuang_2013, S_nchez_2013, S_nchez_2014, S_nchez_2016, Samushia_2014, Anderson_2014a, Anderson_2014b, Beutler_2016a, Beutler_2016b, Zhao_2016, Alam_2017}.  

An alternative and rather more promising tracer is the neutral atomic hydrogen (HI) which pervades the Universe from the recombination epoch through the Cosmic reionization to the present time. During reionization, the intergalactic medium (IGM) was ionized by the first sources. And post-reionization, neutral hydrogen exists only within clouds massive enough to shield themselves from ionizing ultra-violet (UV) photons. These structures are observed as Lyman-alpha absorbers. The stellar and galaxy evolution impacts the distribution of HI in these systems, and therefore, the detection of HI can provide much-needed insights of the galaxy and stellar evolution processes. The cosmic HI can be detected with line emission at the 21cm which arises due to the spin-flip transition of the electron in the atomic hydrogen ground state. The typical temperature of HI in the post reionization epoch ranges upto thousands of Kelvin, which is higher than that of the temperature between the hyperfine states responsible for the 21cm transition. Therefore, the 21cm line transition occurs as emission, which falls within the frequency coverage of many radio telescopes, e.g. MeerKAT \citep{MeerKAT}, ASKAP \citep{ASKAP}, SKA \citep{SKA}. Furthermore, the measured redshift of the HI emission line provides an additional measure of cosmic distance. Thus, it is possible to construct a three-dimensional HI field and therefore measure the fluctuation in the underlying matter distribution. However, the inherent weakness of this signal along with the limited bandwidth of previous telescopes has restricted the detection of HI in individual galaxies to the local Universe ($z\sim 0.1$). 

Fortunately, with the Intensity mapping (IM) technique,  one can construct a low angular resolution 21cm map where the emission from many unresolved galaxies is combined into a single resolution element, boosting the signal \citep{Bharadwaj_Sethi_2001, Wyithe_Loeb_2008, Bull_2015, Santos_2015, Santos_2017}. This approach is analogous to a Cosmic Microwave Background (CMB) map, without the need to detect individual galaxies. The 21cm signal is intrinsically weak compared to the various astrophysical foregrounds which are a few orders of magnitude stronger. Observations of very long duration are required to achieve the required sensitivity with the added complication of maintaining system stability for such long periods. The first tentative detection of the 21cm intensity mapping signal at $z\approx 0.8$ was reported from the Green Bank Telescope (GBT) observations. The cross-correlation signal of GBT observations with DEEP2 optical galaxy survey was first detected by \citet{Chang_2010}; whereas \citet{Masui_2013} reported the cross-correlation signal with the WiggleZ Dark energy survey. In a first-ever attempt in auto-correlation, \citet{Switzer_2013} used the 21cm intensity fluctuation auto-power spectrum to constrain the neutral hydrogen fluctuation at $z\approx 0.8$. 

The purpose of this paper is to demonstrate a complementary approach to single-dish IM experiments, capable of the statistical detection of HI field and therefore the fluctuations in the underlying matter field by measuring the HI power spectrum with interferometric observations. Interferometers have inherent advantages over single-dish measurements. Besides providing high angular resolutions, they are less sensitive to systematics which poses a major problem to the auto-correlation power. However, the smallest $k$-modes accessible to an interferometer is determined by the shortest baselines which may hinder probing the BAO scales. Interferometers such as CHIME \citep{Bandura_2014}, TIANLAI \citep{Xu_2014} and HIRAX \citep{HIRAX} are custom designed to probe the BAO scales using the 21cm signal in the redshift range $z\sim 0.5-2$. In this paper, we study the feasibility of detecting the cosmological 21cm signal with MeerKAT and present forecasts on the statistical measurement of HI with MeerKAT L-band ($856 < \nu < 1712$ MHz) observations on quasi-linear scales. We present the sensitivity estimates at $z\sim 0.27$ from our newly developed simulation pipeline, which is our first attempt towards the measurement of HI power spectrum with MeerKAT. Our pipeline is based on the methods being developed for similar statistical measurement of HI from the Epoch of Reionization from a series of experiments at lower radio frequencies such as LOFAR \citep{Haarlem_2013}, GMRT \citep{Paciga_2013}, PAPER \citep{Parsons_2014}, HERA \citep{DeBoer_2017} and MWA \citep{tingay_2013}. We show that with MIGHTEE \citep{Jarvis_2016, Maddox_2020}, one of MeerKAT's large survey projects, we can achieve constraints on the HI power spectrum at $z=0.27$. There are implicit advantages with such survey projects with a specific emission line. First, it provides a one to one correspondence between observed frequency and redshift, thereby delivering a very high redshift resolution. Secondly, these are generally less time consuming compared to an optical spectroscopic galaxy survey which requires very high sensitivity to detect individual galaxies. 

The paper is structured as follows. In the next section, we provide a theoretical overview of the statistical detection of HI signal and how interferometer measurements enable us to estimate the HI power spectrum. In \autoref{section_simulation}, we give a brief outline of MIGHTEE observations and describe the simulation pipeline to extract the HI power spectrum from MIGHTEE data. The main simulation results, along with the sensitivity estimates, are discussed in detail in \autoref{section_results}. In \autoref{section_forecast}, we forecast the possibility of obtaining constraints on the HI power spectrum with MIGHTEE data and finally, \autoref{section_conclusion} contains the conclusion and scopes of future work. Throughout this paper, we have used the flat $\Lambda$CDM cosmological parameters $[\Omega_m, \Omega_b, h, n_s, \sigma_8] = [0.311, 0.049, 0.677, 0.967, 0.8102]$ from \citet{Planck:2018}.     

\section{Statistical Measurement of HI}
\label{section_HI}
In this section, we formulate the basis for the HI power spectrum analysis through statistical measurements. Although a power spectrum lacks visual representation of the 21cm field like an image; there are some inherent advantages in the statistical approach where we can take the advantage of the Universe being statistically isotropic. Therefore, we can in principle coherently combine the various Fourier modes of the same amplitude although different in direction - which in turn aids in improving the sensitivity. Below, we define the power spectrum following the construction of the 3d Fourier transformation of sky temperature. The sky temperature can be decomposed as: $T({\bm{\theta}}, \nu) = \bar{T}(\nu)[1 + \Delta T({\bm{\theta}}, \nu)]$; where $\bar{T}(\nu)$ and $\Delta T({\bm{\theta}}, \nu)$ are the isotropic and fluctuating component of the temperature distribution; ${\bm{\theta}}$ and $\nu$ denote the position vector on the sky plane and frequency of observation respectively. We define the Fourier transform of the fluctuating component as:
\begin{equation}
  \Delta T({\bm{k}}) = \int_{-\infty}^\infty d^3 r \Delta T(\bm{r})e^{-i \bm{k}\cdot \bm{r}} ,
\end{equation}
where $\bm{r} = \{\bm{\theta}, r_\nu \}$ specifies the 3D position of the emission, $r_\nu$ being the comoving distance to the point of observation and $\bm{k}$ the comoving wave vector. With interferometers, one seeks to compute the two-point correlations of the cosmological signal and the most significant correlation function is the power spectrum $P(k)$, defined as:
\begin{equation}
    \langle \Delta T^*({\bm{k}}) \Delta T({\bm{k'}}) = (2\pi)^3 \delta^3(\bm{k}-\bm{k'})P(k) .
\end{equation}

Radio interferometers calculate the spatial correlation of electric fields from the sky with the measured visibility, obtained by correlating data from each antenna pair. Under the flat-sky approximation, the visibility can be expressed as:
\begin{equation}
    V(\bm{b}, \nu) = \int A(\bm{\theta}, \nu) \Delta T({\bm{\theta}}, \nu) e^{-i2\pi\nu \bm{b}\cdot\bm{\theta}/c} d\Omega.
\end{equation}
Here, $\bm{\theta}$ refers to the position on the sky, $A(\bm{\theta}, \nu)$ is the primary beam response of the telescope, $\bm{b}$ denotes the baseline vector in physical units corresponding to each antenna pair and $d\Omega$ being the solid angle element. The cosmological HI power spectrum can be estimated from measured visibilities in the form of `delay spectrum' by the following relation \citep{Morales_2004, Mcquinn_2006, Parsons_2012a, Parsons_2014, Liu_2020}:
\begin{equation}
P_{\rm{D}}(\bm{k_\perp}, k_\parallel) \equiv \frac{A_e}{\lambda^2 B} \frac{x^2 y}{B} |V(\bm{b},\tau)|^2 \left(\frac{\lambda^2}{2k_B}\right)^2.
\label{P}
\end{equation}
Here, $A_e$ and B are the effective antenna area and bandwidth respectively, $\lambda$ is the wavelength at the centre of the band, $k_B$ is the Boltzmann constant, $x$ denotes the comoving distance to the redshift $z$ corresponding to $\lambda$, whereas $y$ signifies the comoving width along the redshift axis corresponding to B. In \autoref{P}, we have decomposed the wave vector $\bm{k}$ into the components on the plane of the sky $\bm{k_\perp}$ and along the line of sight $k_\parallel$; and they are related to the interferometric variables as:
\begin{equation}
\bm{k_\perp} = \frac{2\pi \bm{b}}{\lambda x} ; \hspace{10pt} k_\parallel = \frac{2\pi\tau\nu_{21}H_0 E(z)}{c(1+z)^2}
\label{conversion}
\end{equation}
where, $\nu_{21}$ is the rest frame frequency of the 21cm line; $H_0$ and $E(z)=[\Omega_M(1+z)^3 + \Omega_K(1+z)^2 + \Omega_\Lambda]^{1/2}$ are the standard cosmological parameters. $V(\bm{b},\tau)$ is the Visibility function in the delay space ($\tau = \bm{b}\cdot\hat{s}/c$) obtained by delay transforming the measured visibilities with an FFT. The dissimilarity in spectral behaviour between the HI signal and foregrounds makes it possible to isolate the latter in the Fourier space. The `delay space' approach \citep{Parsons_2012a, Parsons_2012b, Vedantham_2012, Liu_2014a, Liu_2014b, Paul_2016} takes advantage of this property.
The foregrounds are smooth in the frequency domain as they originate from continuum emissions such as the Synchrotron emission from both our galaxy and other extragalactic sources. Therefore in the Fourier space, they are restricted to fewer Fourier modes. The HI signal, on the other hand, has different characteristics in the frequency domain as the frequency is a measure of cosmological distance and therefore has significant structures in the frequency space. This feature potentially allows us to separate the HI from foregrounds. In this approach, the visibilities observed by each antenna pair are Fourier transformed along the frequency axis, which isolates the foreground contribution in the `delay space'. The Fourier conjugate variable can be associated with the line of sight cosmological distance, and the `delay spectrum' constructed from this method is capable of recovering the cosmological 3d HI power spectrum. One caveat in this approach is that the Fourier mode on sky plane ($\bm{k_\perp}$) is calculated at the centre of the frequency band for each baseline. However, in the actual scenario, each physical baseline corresponds to a range of $\bm{k_\perp}$ modes across the bandwidth. The span of the $\bm{k_\perp}$ modes is higher with increasing baseline length and bandwidth. However, if one restricts to shorter baselines (where the sensitivity is higher due to a large number of uv points) and small frequency range, this effect is not severe and the `delay power spectra' is a good approximation to the actual cosmological 3d HI power spectrum \citep{Parsons_2012b, Liu_2014a}. Also, this approach comes with an added advantage that one can work with the data from individual baselines (which is the regular format of the primary data output of a radio interferometer, i.e. visibilities). 

\autoref{P} assumes that the change in $\bm{k_\perp}$ is minimal across the bandwidth for baselines considered for power spectrum calculation to justify the conversion from $\tau$ to $k_\parallel$ in \autoref{conversion} \citep{Liu_2014a}. This approximation holds well for short baselines and small frequency range over which the delay transform is performed. We estimate that at $k \sim 10 \ {\rm Mpc}^{-1}$, an error of approximately $2\%$ is introduced on the frequency dependent $\bm{k_\perp}$ in this approach.

\subsection{HI signal}
\label{HI_model}
The prime observable in the HI intensity mapping experiments is the 21cm emission line from neutral hydrogen. The mean brightness temperature of the HI 21-cm emission can be expressed as \citep{Santos_2015,Santos_2017}:
\begin{equation}
   \bar{T_b}(z) \approx 566h \left(\frac{H_0}{H(z)}\right) \left(\frac{\Omega_{\rm{HI}}(z)}{0.003}\right)(1+z)^2 \mu \rm{K}.
\end{equation}
Here, $H(z) = H_0 E(z)$ and $\Omega_{\rm{HI}}(z)$ is the neutral hydrogen density function:
\begin{equation}
    \Omega_{\rm{HI}}(z) = \frac{\rho_{\rm{HI}}(z)}{\rho_{c,0}(1+z)^3},
\end{equation}
where $\rho_{\rm{HI}}(z)$ and $\rho_{c,0}$ are the proper HI density and critical density of the Universe at $z=0$ respectively. $\Omega_{\rm{HI}}(z)$ is a crucial quantity in determining the hydrogen content of the Universe at various redshifts and therefore plays a significant role in the calculation of the 21-cm brightness temperature. Several experiments have measured $\Omega_{\rm{HI}}(z)$ over a range of redshifts. Direct 21-cm observations from galaxies have measured this quantity at low redshifts \citep{Zwaan_2005a, Jones_2018}; whereas the quasar absorption spectra in the damped Ly$\alpha$ systems have put constraints on $\Omega_{\rm{HI}}(z)$ at higher redshifts ($z > 2$) \citep[e.g.,][]{Rau_2006, Prochaska_2009, Noterdaeme_2012, Font-Ribera_2012, Zafar_2013, Crighton_2015, Neeleman_2016, Sanchez_2016, Bird_2017}. The HI spectral stacking has been used to constrain the HI abundance at the intermediate redshift range $0.2 < z < 2$ \citep{Lah_2007, Delhaize_2013, Rhee_2013, Rhee_2016, Kanekar_2016, Rhee_2018}; and HI emission studies with ASKAP, MeerKAT and SKA are expected to explore this range in more detail. 

The HI signal follows the underlying dark matter fluctuation and therefore the brightness temperature as a function of position and frequency is given by:
\begin{equation}
    T_b(\nu,\Omega)\approx \bar{T}_b(z)\left[1 + b_{\rm HI}\delta_m(z)-\frac{1}{H(z)}\frac{dv}{dr_\nu}\right],
\end{equation}
where $v$ is the peculiar velocity of emitters. The HI density function $\rho_{\rm{HI}}(z)$ and bias function $b_{\rm HI}(z)$ can be computed using the halo mass function ($\frac{dn}{dM}$) and the HI mass content inside a dark matter halo of mass $M$, $M_{\rm HI}$:
\begin{equation}
\rho_{\rm HI}(z) = \int_{M_{\rm min}}^{M_{\rm max}}dM \frac{dn}{dM}(M,z) M_{\rm HI}(M,z),
\end{equation}

\begin{equation}
b_{\rm HI}(z) = \frac{1}{\rho_{\rm HI}}\int_{M_{\rm min}}^{M_{\rm max}}dM \frac{dn}{dM}(M,z) M_{\rm HI}(M,z)b(M,z),
\end{equation}
where $b(M,z)$ is the halo bias.

In this paper, we assume a simple power-law model of the halo mass following the prescription of \citet{Santos_2015}: $M_{\rm HI}(M)=AM^\alpha$ with $\alpha=0.6$ and $A\sim 220$ that fits both low and high redshift observations within reasonable accuracy. The scaling relations for all relevant quantities to compute the HI signal are obtained with the above formulation and are used throughout this paper (see \citet{Santos_2017} for details). With all these parameters in place, the HI power spectrum in redshift space can be computed in terms of the matter power spectrum $P_{\rm M}(k,z)$ and the bias function as:
\begin{equation}
    P_{\rm HI}(k,z) = \bar{T}_b(z)^2 b_{\rm HI}(z)^2 P_{\rm M}(k,z).
    \label{PHI}
\end{equation}

\section{ sensitivity for estimating HI power spectrum}
\label{section_simulation}
\citet{Santos:2015dsa} showed that SKA1-Mid will have the required sensitivity for a reasonable amount of integration time to constrain the cosmological parameters; however precursor telescopes like MeerKAT should be able to integrate down to such sensitivities on deep single pointings. The MeerKAT radio telescope is located in the Karoo region of South Africa. The telescope array consists of 64 dish antennas of $13.5$ meter diameter. The central core region of 1 km diameter houses 48 antennas, whereas the other 16 antennas are distributed up to a radius of 4 km from the centre. The dense core of MeerKAT facilitates higher sensitivity at low $\bm{k}_\perp$ modes which can aid the statistical detection of HI at relevant cosmological scales using the interferometer data.

The simulation pipeline outlined in this paper aims to present realistic outcomes that can be compared with the real data. Considering that, the pipeline needs to incorporate contributions that are present in the real data, which includes HI, noise and foregrounds. Along with an input HI model (described in \autoref{section_HI}), the thermal noise can be modelled as random processes with the help of various system parameters. For foreground modelling, we choose to adopt a discrete point source model which we create by imaging the MIGHTEE field of interest as described in the following subsection.

\subsection{MIGHTEE}
In this paper, we use data from MIGHTEE survey for the sensitivity estimation. In particular, we process the single pointing COSMOS field observation with an on-source integration time of $\sim 11.2$ hours. The observation took place with the 64-antenna MeerKAT configuration spanning over two days -- 19-Apr-2018 and 06-May-2018. \citet{Maddox_2020} provides a detailed description of the MIGHTEE survey design and observational methods. Along with the target COSMOS field, the fields J0408-6545 and 3C237 were observed as the primary and secondary calibrators respectively. To estimate the Noise power spectrum and therefore, the sensitivity level, one only requires the uv distribution and telescope information such as system temperature, effective area, time and frequency resolution. Also, in theory, one can compute the delay power spectrum from the calibrated visibility data itself without going to the image domain, assuming that foreground isolation is reasonably accurate in the final power spectrum. However, we perform flagging, calibration on the raw data with the purpose-built processMeerKAT\footnote{The processMeerKAT pipeline has been developed and maintained by the Inter-University Institute for Data Intensive Astronomy (IDIA) Pipelines team. For further details see \url{https://idia-pipelines.github.io/docs/processMeerKAT}} pipeline for MeerKAT data calibration; and initial processing in the image domain to obtain a point source model to replicate the foreground contribution in our simulation pipeline. The processMeerKAT uses CASA \citep{CASA} based algorithms to flag RFI contaminated components and bad data; compute phase and flux gains from the reference calibrator observations. The primary calibrator J0408-6545 was used by processMeerKAT to estimate the delay and bandpass solutions which were further applied on the secondary calibrator 3C237 to calculate the time-dependent complex gains. All the gain corrections were then applied to the target COSMOS data. The raw data had an integration period of 4 seconds and spanned a total 4096 channels of 208.984 kHz channel-width in L-band. Post flagging and calibration, the data was split to a sub-band of $950 - 1150$ MHz and time-averaged to 8 seconds, which was further processed for continuum imaging. As we are not using the full available band for the continuum imaging, the resulting model will lack contributions from fainter sources. However, as we will see later, the detected point sources in this band are good enough for an accurate model of the foreground contamination. 

\begin{figure*}
\centering
\includegraphics[trim={1cm 10.5cm 1cm 2cm},clip,width=1.0\textwidth]{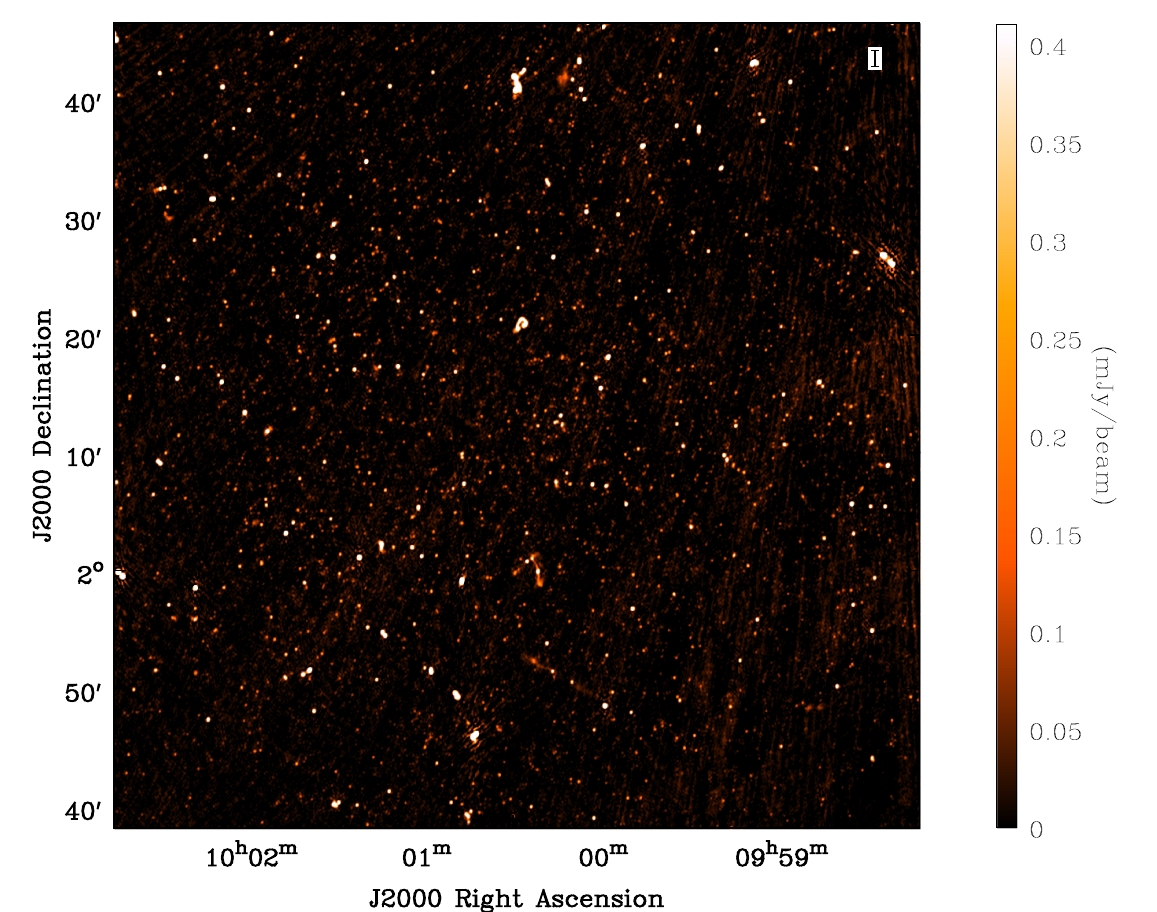}
\caption{Frequency-averaged Stokes I image of the COSMOS field at 1115.14 MHz from 11.2 hours of tracking sub-band data ($950 - 1150$ MHz) with the map-rms $\sim 10 \mu$Jy beam$^{-1}$.  The discrete point source model generated in the imaging process is further used as the foreground model in our simulation pipeline.}
\label{COSMOS}
\end{figure*}

After flagging and calibration by the processMeerKAT pipeline, the following steps are used for the continuum imaging process. We apply the CASA \textit{tclean} and \textit{gaincal} tasks on the MS file for deconvolution and self-calibration respectively. We have used the multi-scale multi-frequency (MS-MFS) synthesis algorithm \citep{Rau_2011} for the imaging with \textit{nterms=2} to estimate the intensity along with its spectral variation. To account for the non-coplanarity of the MeerKAT baselines, we also used the W-projection algorithm \citep{Cornwell_2008}. To reduce the error emanating from the temporal variations in the system gain, we perform a few rounds of self-calibration with \textit{gaincal} (3 rounds phase and 1 amplitude+phase) and \textit{tclean} loop. For the first few iterations, the number of clean iterations is set to a low number to avoid cleaning deep and detect false source components from noise pixels. With each self-calibration, the number of clean iterations is increased gradually. At the final \textit{tclean} task, the threshold for the clean is set at $5\sigma$ level where $\sigma$ is the standard deviation in the residual image. We create an image of size $1024 \times 1024$ pixels with 8 arcseconds cell size. The image size has been kept larger compared to the main field of view to include the bright sources from the sidelobes. We do not perform any primary beam correction to the image. This is actually an advantage as it means that primary beam effects will be included in our point source foreground model. In \autoref{COSMOS}, we present the central region of the image of size $1.14^{\circ} \times 1.14^{\circ}$. From the $11.2$ hours of sub-band data, we obtain an rms $\sim 10 \mu$Jy beam$^{-1}$ from the resulting residual image. To check the calibration accuracy, we have compared the flux scales of some known sources in the final image and they are consistent with the historical values. The model obtained from the final imaging step is further used as the foreground model in the power spectrum pipeline as described in the following section. As the foreground model is generated directly from the CLEAN components, the effects of primary beam are implicitly included in the model. 

\subsection{Simulation Pipeline}
Each visibility measurement by the interferometer receives a contribution from system noise. This, along with the cosmological signal itself, adds to the uncertainty in $\bm{k}$ space. Therefore, it is important to have a measure of thermal noise contribution to estimate the power spectrum sensitivity level. In this section, we describe the details of our simulation pipeline. For this part, we use only a small subset of the available data by splitting the data further on a narrower band of $B = 220 \times 208.984$ kHz centred at 1115.14 MHz which corresponds to $z=0.27$. Below, we delineate the steps of our simulation pipeline in detail:

\begin{figure*}
\centering
\includegraphics[width=1.0\textwidth]{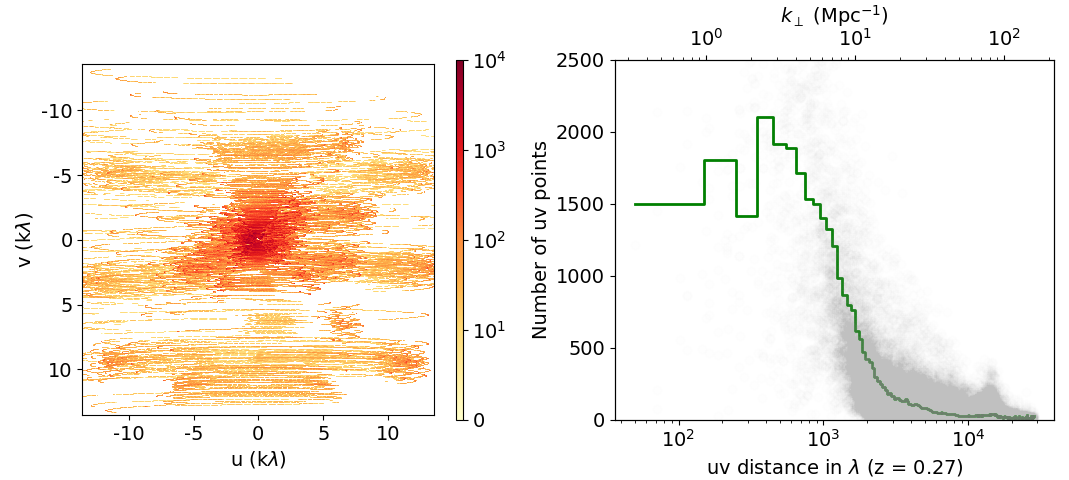}
\caption{Left panel: Distribution of baselines on a 2d uv plane for 11.2 hours tracking of the COSMOS field. The uv plane is segmented onto a discrete grid with cell-size $\Delta u = \Delta v = 60\lambda$. The color represents the number of $uv$ points on the grid, which is maximum at the centre and falls with increasing baseline length. Right panel: Baseline population as a function of $uv$ distance. The solid line represents the average number of baselines as a function of $uv$ distance (bin size: $\Delta uv = 100\lambda$). Also, each cell is plotted as a function of its $uv$ distance on the lower x-axis, whereas the upper x-axis shows the corresponding $k_\perp$ estimates. The y-axis denotes the number of $uv$ points. The number is high at small $uv$ distance, implying increased sensitivity at low $k_\perp$ and it falls with increasing $k_\perp$.}
\label{uv}
\end{figure*}

\begin{enumerate}
    \item  Each physical baseline corresponds to a (u,v) coordinate which changes after every integration interval ($t_{\rm{int}} = 8$  seconds) over the period of tracking. The uv coordinates are extracted for the entire duration of the data ($t_{\rm{total}} \sim 11.2$ hours) from the visibility file. Note that the (u,v) points are calculated at the band center. Before flagging, its number should be: $N_{uv} = \frac{N_{\rm{ant}}(N_{\rm{ant}}-1)}{2} \frac{t_{\rm{total}}}{t_{\rm{int}}}$; where $N_{\rm{ant}}$ is the number of antennas used in the observation. However, we do not include the flagged baselines in our simulation. Each baseline receives contributions from the HI, foreground and noise. Therefore, for the \textit{i}'th baseline, the visibility can be expressed as: $V_i = V_{{\rm{HI}},i} + V_{{\rm{FG}},i} + V_{{\rm{N}},i}$. $V_{\rm{HI}}$ is the contribution from an input model HI signal which we want to recover from the final power spectrum; whereas $V_{\rm{FG}}$ and $V_{\rm{N}}$ are the foreground and noise components respectively.
    
    \item In \autoref{COSMOS}, we observe that the discrete extragalactic sources dominate the radio sky in the frequency range of our interest. Other than bright radio sources, the Galactic synchrotron emission originated from the interactions of cosmic-ray electrons and the interstellar magnetic field, is expected to contribute to the total foreground budget. However, this diffuse emission has a strong presence along the Galactic plane. As the COSMOS target field is far from the Galactic disc, the contribution from the Galactic synchrotron emission is significantly low, and we do not see any considerable diffuse structure in the continuum image (\autoref{COSMOS}). This suggests that the diffuse Galactic component has a weak contribution on these angular scales, probably only detectable after subtracting the extragalactic point sources. Therefore in our foreground model, we only consider the bright extragalactic sources which are extracted as \textit{CLEAN} components of the continuum image during the deconvolution process. 
    Application of the source detection algorithm PyBDSF \citep{Mohan_2015} on the continuum image reveals a total of 3391 extragalactic discrete point sources ($\sim$ mJy radio flux density) in our foreground model.   
    
    \item For the system noise contribution, we generate $V_{\rm{N}}$ per baseline, which we model as a Gaussian random variable. Therefore at each baseline and channel, the real and imaginary part of $V_{\rm{N}}$ are generated from a random process with rms calculated by the following relation \citep{Taylor_1999} which provides the thermal noise estimate per baseline:
    \begin{equation}
        \sigma_{\rm{N}} = \frac{2k_{\rm{B}} T_{\rm{sys}}}{A_e\sqrt{\delta\nu \delta t}}.
        \label{sigma_N}
    \end{equation}
    Here, $T_{\rm{sys}}$ and $A_e$ are the system temperature and effective area of each antenna respectively, $k_{\rm{B}}$ is the Boltzmann constant, $\delta\nu$ the channel width and $\delta t$ the time resolution. The corresponding values used to calculate $\sigma_{\rm{N}}$ are: $\delta\nu = 208.984$ kHz, $\delta t = 8$ sec and  $A_e/T_{\rm{sys}} = 6.22$ m$^2$/K \footnote{MeerKAT array specification document as on 2016-10-10, available in: \url{https://www.sarao.ac.za/science/meerkat/about-meerkat/}}. The contribution of system noise is approximately constant across the frequency channels (and therefore along $k_\parallel$); it only varies across $\bm{k_\perp}$ as the noise level depends on the baseline density (\autoref{uv}).
    
    \item Next, we generate a $uv\nu$ cube by segmenting the uv plane onto a discrete grid with cell-size $\Delta u = \Delta v = 60 \lambda$; this choice of grid resolution is motivated by the primary beam size in the Fourier domain. The third axis of the $uv\nu$ cube is in frequency which is already discretized into channels of width $208.984$ kHz. In \autoref{uv}, we show the distribution of uv points. On the left panel of \autoref{uv}, each data point corresponds to a uv pixel, and the color shows the number of uv points within every cell. On the right panel, the mean number of uv points is plotted as a function of $uv$ distance from the origin ($u=v=0$). On smaller $uv$ distance, the uv points are densely populated; and as $|\bm{k_\perp}|$ is proportional to the uv distance, it translates to higher sensitivity at small $\bm{k_\perp}$ modes. It is worth mentioning here that the calibrated visibility measurements have some channels flagged due to excessive RFI presence. These gaps are reflected in the foreground model visibility as well. If we consider visibilities filled with too many flagged channels for the power spectrum calculation, it will cause a spurious flow of foreground power to higher $k_\parallel$ modes. Therefore, we only consider those $uv$ points for which the visibility measurements have at least $80\%$ unflagged channels (out of 220). Moreover, to minimize the contamination from the remaining flagged channels, we substitute each flagged foreground component with the foreground visibility from the nearest neighbour unflagged channel. This is to make sure that the simulated visibilities have no channels with \textit{zeros} while performing the delay transformation along the frequency axis. For the real data, the same selection rule applies for assigning the baselines on the grid; and the flagged channels are replaced with the visibility entry from the nearest neighbour unflagged data channel.

    \item The visibilities within a uv pixel are then averaged assuming the sky signal to be the same across all baselines contributing to that grid point. At this stage, we also add the contribution from the input model cosmological signal $V_{\rm{HI}}$ to the averaged visibility per grid point, which we generate as a random process with variance calculated from \autoref{PHI}. This ensures that the resulting `delay spectrum' (\autoref{P}), estimated from the complex HI visibilities, agrees with the input HI power spectrum. Next, we perform the delay transformation with an FFT along the frequency axis. During the FFT, each visibility per grid-point is multiplied with a spectral weighting function (Blackman-Harris Window) to suppress the leakage of foreground power to higher $k_\parallel$ modes. The resulting $V(u,v,\tau)$ are then used in \autoref{P} to compute the 3d power spectrum in $(\bm{k_\perp}, k_\parallel)$ domain following the conversions listed in \autoref{conversion} where $(u,v)$ are calculated at the grid-center. The power estimates $P(\bm{k_\perp}, k_\parallel)$ in 3d $\bm{k}$ space are further combined with an inverse variance (thermal noise) weighting to compute the 2d and 1d power spectrum $P(k_\perp, k_\parallel)$ and $P(k)$ respectively, where $k_\perp = |\bm{k_\perp}|$. The lowest $k_\perp$ mode probed by the $uv$ distribution is indicated by the smallest baseline; and for the case in hand, ${(k_\perp)_{\rm{min}}} \sim 0.33$ Mpc$^{-1}$. Therefore, we choose the bin width $\Delta {k_\perp} = 0.35$ Mpc$^{-1}$ for estimating $P(k_\perp, k_\parallel)$. The lowest mode along $k_\parallel$ and the bin width follows the bandwidth $B$ used for FFT, and is given by $\Delta k_\parallel \approx 0.031$ Mpc$^{-1}$. To compute the 1d power spectrum $P(k)$, we use a logarithmic binning across $k$, i.e. the bin size gets larger with increasing $k$; and the smallest bin width is $\Delta k \sim 0.4$ Mpc$^{-1}$.
\end{enumerate}

\section{Results}
\label{section_results}
\begin{figure*}
\subfloat[HI]{\includegraphics[width=0.5\textwidth]{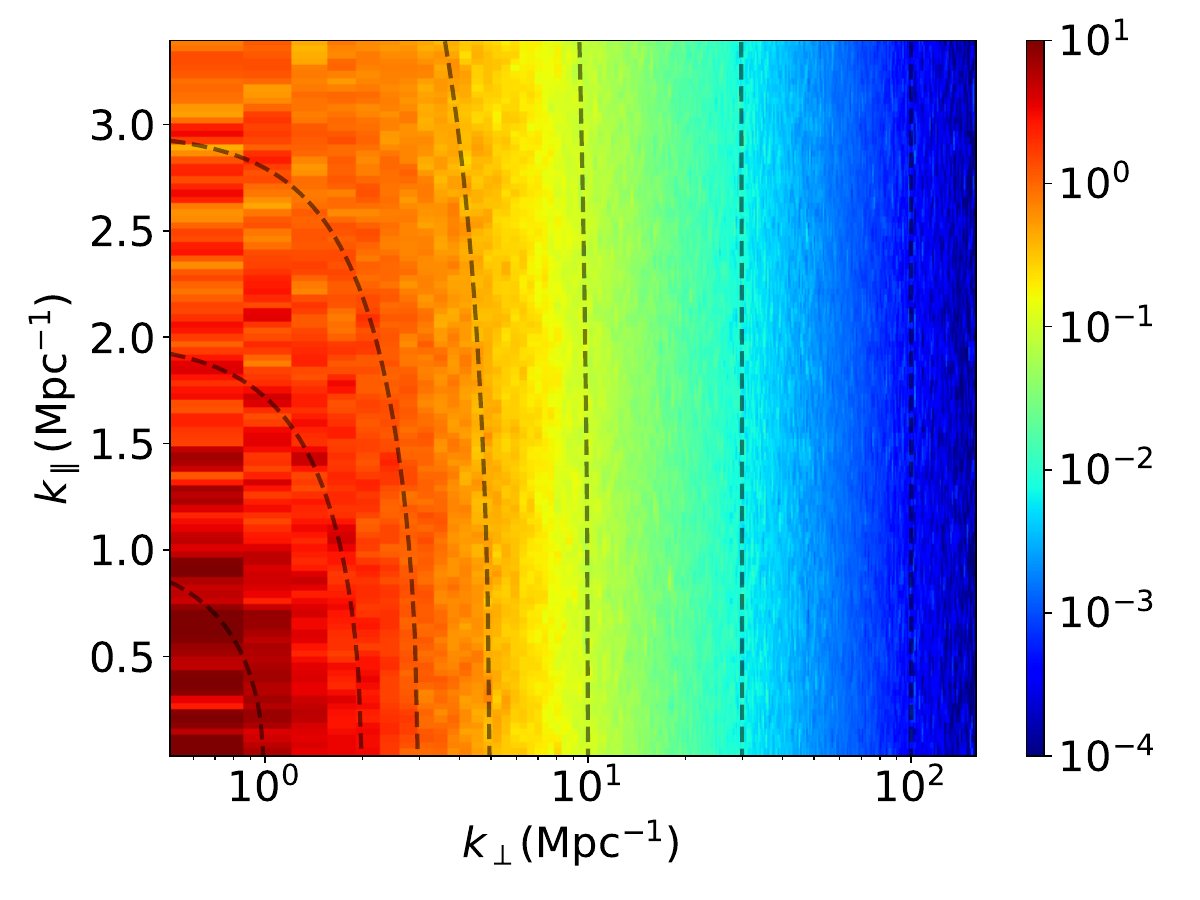}}
\subfloat[HI + TN]{\includegraphics[width=0.5\textwidth]{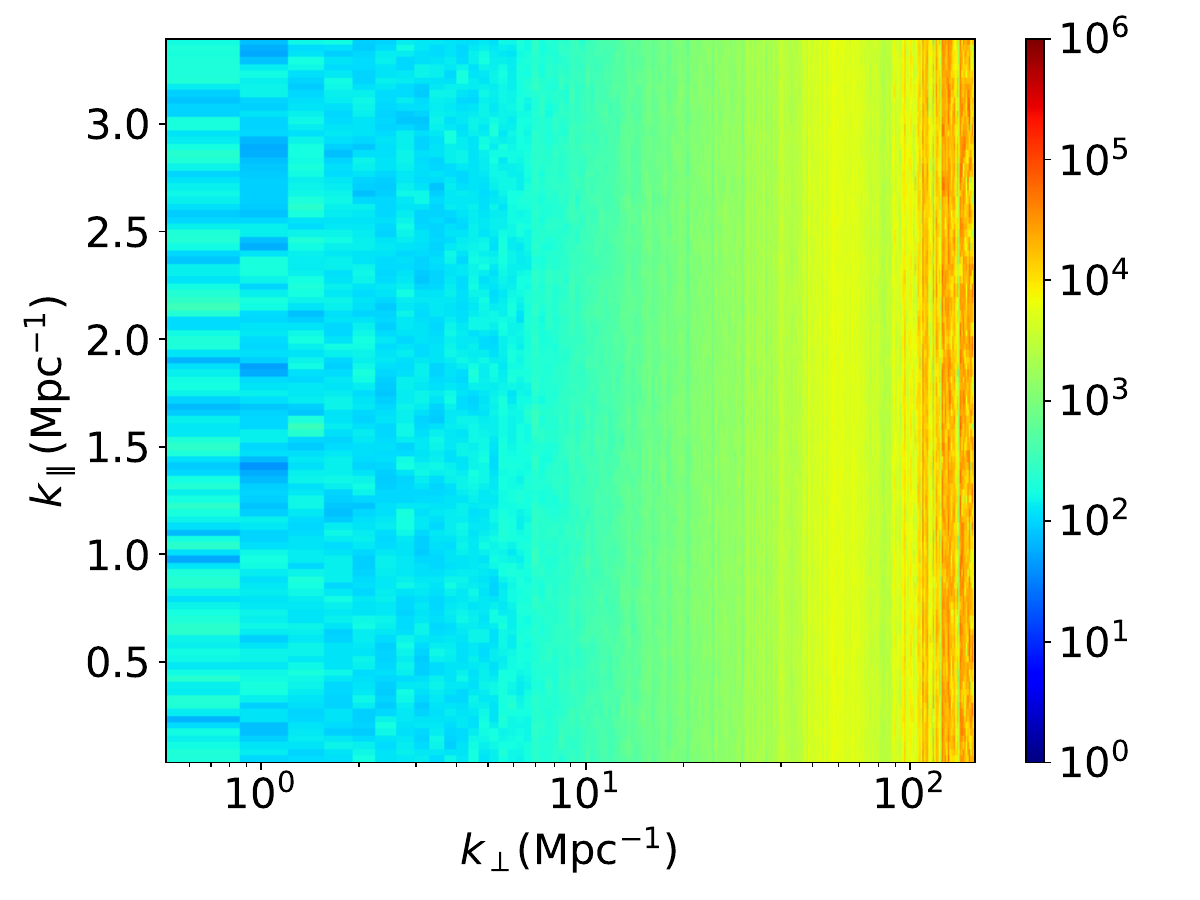}}
\caption{2d power spectrum in the unit of $\rm{mK}^2 \rm{Mpc}^3$ generated from the simulation pipeline for a single realization, the dashed lines denote contours of constant radius $k$. (a) shows the case for only HI; whereas in (b), the contribution from both HI and thermal noise are considered.}
\label{TNHI_2d}
\end{figure*}

\begin{figure*}
\centering
\subfloat[Full Simulation]{\includegraphics[width=0.5\textwidth]{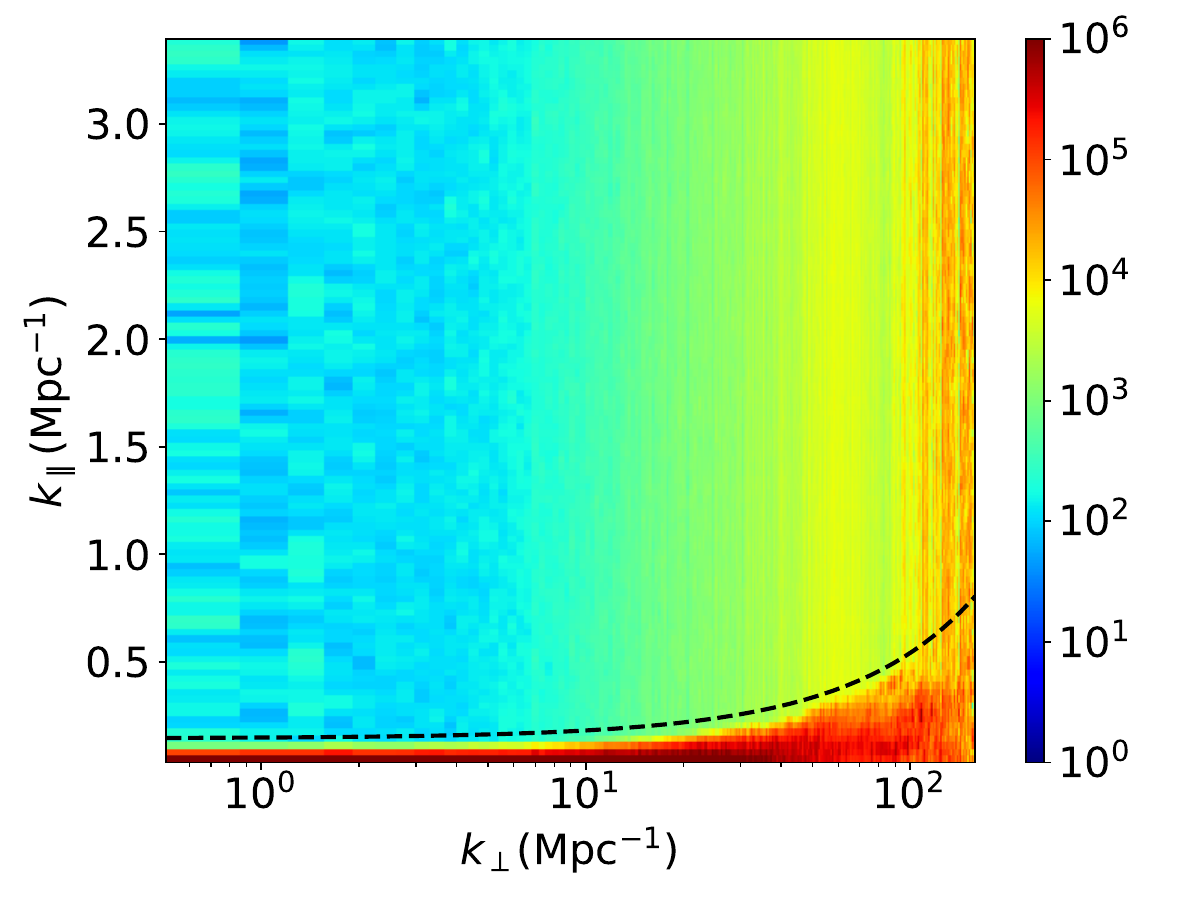}}
\subfloat[Calibrated data (Stokes I)]{\includegraphics[width=0.5\textwidth]{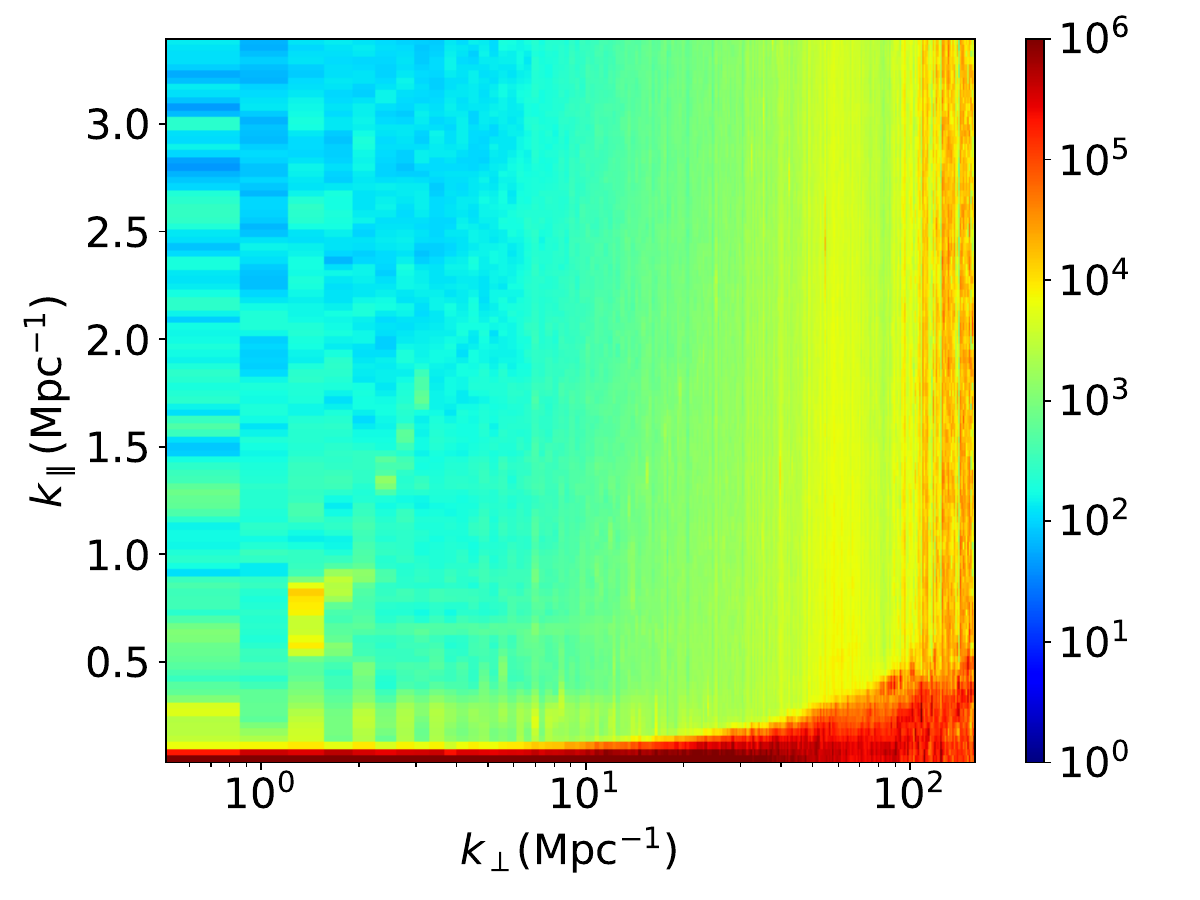}}
\caption{(a) Simulated 2d power spectrum (single realization) for 11.2 hours tracking of the COSMOS field, including contributions from the model cosmological signal, thermal noise and model foreground (\autoref{COSMOS}). The foreground contribution is well isolated at lower $k_\parallel$ values and the region beyond the foreground wedge is dominated by noise and the 21cm signal. The dashed black line denotes the boundary of the foreground wedge (\autoref{wedge_eq}) and the pixels above the line are used to compute the 1d power spectra. (b) 2d power spectrum generated from the calibrated visibility data (Stokes I) using the same pipeline. Both the plots are in qualitative agreement in terms of foreground isolation and overall power level (in units of $\rm{mK}^2 \rm{Mpc}^3$).}
\label{Full_2d}
\end{figure*}

\begin{figure*}
\centering
\subfloat[Power ratio between Stokes V and Thermal Noise simulation in 2d]{\includegraphics[width=0.5\textwidth]{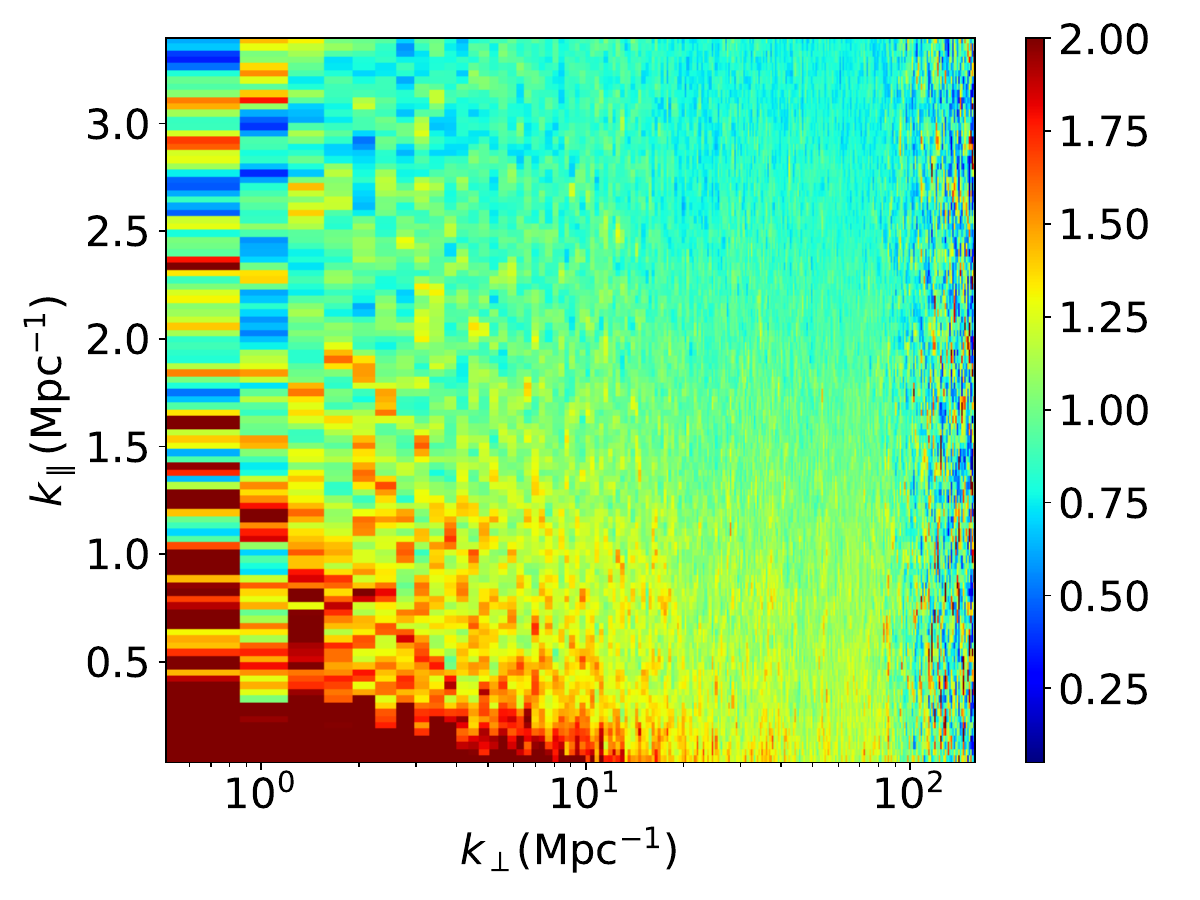}}
\subfloat[Stokes V and Thermal Noise comparison in 1d]{\includegraphics[width=0.5\textwidth]{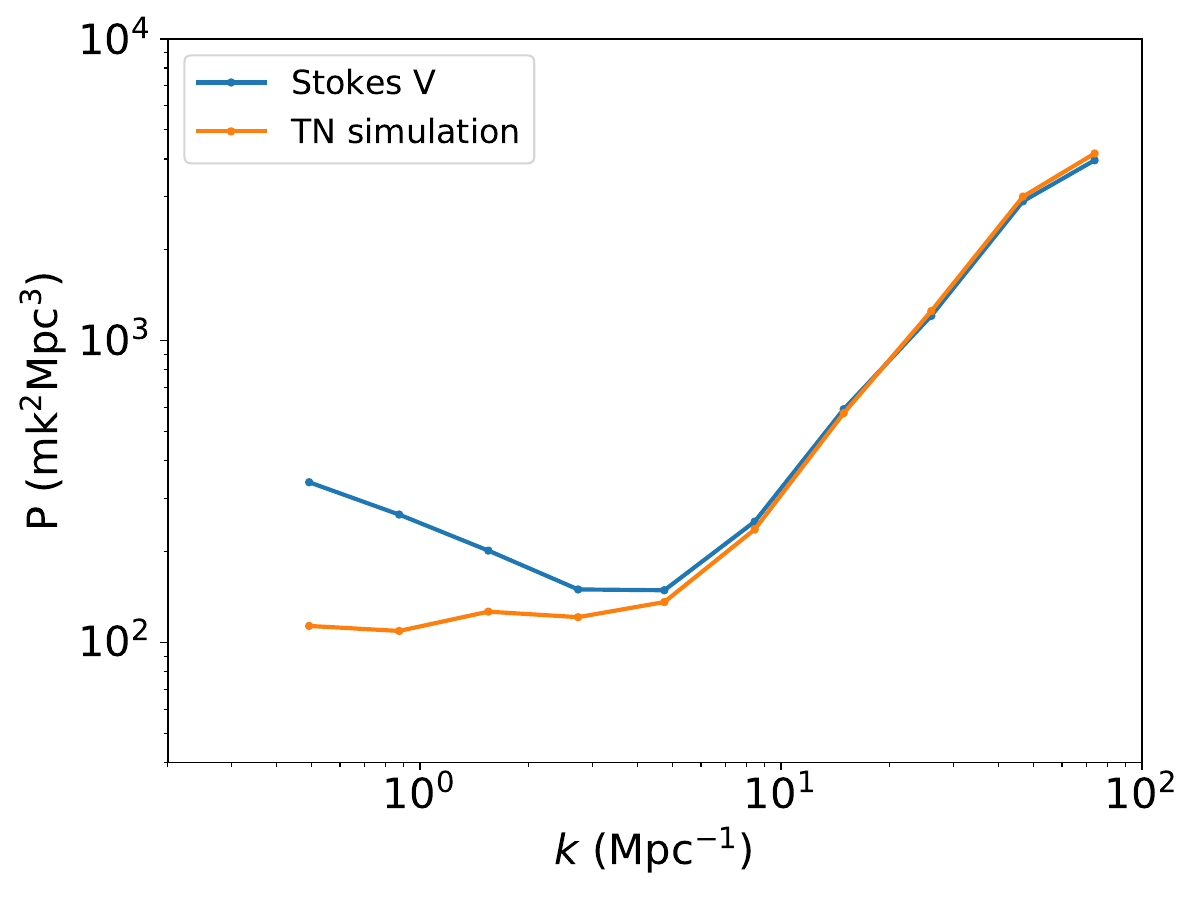}}
\caption{Stokes V from data and simulated thermal noise comparison; the thermal noise model is calculated as the average of 1000 realizations of noise power spectrum estimated from the simulation pipeline where the noise visibility $V_{\rm{N}}$ is used as the only input.}
\label{V_TN}
\end{figure*}

The target 21cm signal can provide new insights into cosmology as well as the properties of low mass HI galaxies which are difficult to detect directly otherwise. However, since the signal is buried underneath strong foregrounds and noise, it is imperative to check the detected signal is indeed cosmological in nature and the processes and techniques used in the measurement are lossless. The prime objective of our simulation pipeline is, therefore, to demonstrate that we can recover a cosmological input signal in the presence of strong foregrounds and system noise. In \autoref{TNHI_2d}a, we present the 2d power spectrum $P_{\rm{HI}} (k_\perp, k_\parallel)$ for a single realization where we have considered only HI and ignored the foreground and noise contributions. This shows that the HI signal is maximum at low $k$ and exhibits an isotropic nature.

The presence of strong foregrounds curtails the observability of the 21cm signal. Two possible approaches can be undertaken to deal with the extragalactic foregrounds. One can try to model the foregrounds with great accuracy and subtract in the image domain leaving behind only the 21cm signal and Gaussian noise. This can be particularly successful with point sources for which we already have information such as position, flux and spectral index. Another approach is to rely on the fact that foregrounds are expected to be smooth in frequency and therefore potentially distinguishable from the cosmological signal that has significant spectral structures. For the 21cm signal, the frequency is a measure of the redshift, and thereby of the line of sight distance to the emitters. Foregrounds do not have any such cosmological significance, and hence in the delay space, the foreground contribution can be restricted to the first few Fourier modes. One can take advantage of this difference and separate the signal using foreground cleaning methods \citep{Alonso_2015, Wolz_2015, Switzer_2015, Wolz_2017} or use the so called foreground avoidance approach \citep{Datta_2010,Morales_2012, Parsons_2012a, Parsons_2012b,Vedantham_2012, Pober_2013, Thyagarajan_2013, Paul_2016}.

The other component, thermal noise, is inversely proportional to baseline density which is maximum at short $uv$ values (for MeerKAT) that correspond to the lowest $k_\perp$ modes. In \autoref{Full_2d}a, we present the 2d power spectrum $P_{\rm{Full}} (k_\perp, k_\parallel)$ for a single realization which includes all three visibility components. The thermal noise contribution is simulated from the $uv$ distribution for the 11.2 hours tracking case of the COSMOS field, whereas we have used the discrete source model of the COSMOS field generated in the imaging process for the foreground component. \autoref{Full_2d} bears out the previous assumption that thermal noise contribution is lowest at small $k_\perp$, and grows with increasing $k_\perp$ value which is in accordance with \autoref{uv}b. Also, the strongest 21cm signal lie at shortest $k_\perp$ modes and decreases rapidly with increasing $k_\perp$ values (\autoref{TNHI_2d}a). As discussed above, the foreground contribution is isolated and occupy a wedge-shaped region due to the smooth spectral characteristics \citep{Datta_2010,Parsons_2012b,Thyagarajan_2013,Liu_2014b, Paul_2016}. The foreground contamination is suppressed in the region beyond the wedge, and this space can be expected to be dominated by the 21cm signal and Gaussian noise. To assess the extent of foreground contamination, we also show the contribution from the HI plus thermal noise only case in \autoref{TNHI_2d}b. 

To compare to the simulation result, we present the 2d power spectrum computed from the calibrated Stokes-I visibility data in \autoref{Full_2d}b using the same pipeline. Both plots in \autoref{Full_2d} show striking similarities; in particular, this comparison allows us to assess the efficacy of the foreground isolation approach, and compare the power levels within and beyond the foreground wedge. It also indicates the range of scales that can be probed with the current setup. However, we do not correct for noise-bias and possible instrumental systematics while calculating the power spectrum in \autoref{Full_2d}b. A comprehensive direction-dependent calibration strategy might be required to mitigate the errors due to sources away from the center of the field and ionospheric effects, but we do not see such effect at these levels. There are some visible features in \autoref{Full_2d}b at low $k_\parallel$ which are possibly due to foreground leakage, systematics or residual RFI. We do not know the exact origin of them yet. We are investigating them in detail and working on possible mitigating strategies. We present \autoref{Full_2d}b as a comparison with the simulation to define the wedge. The sensitivity estimates presented in the following section are fully based on the simulations and the contamination  outside the wedge are not taken into account.

We further compare the noise model in the simulation pipeline to the Stokes V power spectrum. The extragalactic foreground sources have negligible circular polarization, and therefore, the Stokes V mode is a good estimator of the thermal noise in the system. We compute the Stokes V visibilities from the calibrated dataset, and estimate the corresponding 2d and 1d power spectra. In \autoref{V_TN}a, we show the power ratio between Stokes V and our thermal noise model (generated from the simulation pipeline with $V_N$ as the only input). The ratio plot shows excess power in Stokes V at low $k$, which indicates large scale polarization leakage in the calibrated data. At higher $k$, the Stokes V power spectrum exhibits noise-like features, and it is in good agreement with our thermal noise model (\autoref{V_TN}b). Note that, \autoref{Full_2d}b and \autoref{V_TN} are presented to compare the accuracy of our simulation pipeline in estimating the noise power spectrum, we do \textit{not} present any data-results further in this paper. 

Next, we compute the 1d power spectra along shells of constant $\abs{k}$. While doing so, we exclude the  $(\bm{k_\perp}, k_\parallel)$ cells contaminated by the foreground wedge. More precisely, we use the following selection criteria
\begin{equation}
    k_\parallel < \frac{x H_0 E(z) \sin({\theta})}{c(1+z)} k_\perp 
    \label{wedge_eq}
\end{equation}
where $\theta$ refers to the angular extent of the MeerKAT beam. To obtain good statistics, we generate $1000$ realizations from the simulation pipeline. In these cases, we generate different realizations of the noise and 21cm signal with the same foreground model. 

\section{Detectability}
\label{section_forecast}

\begin{figure*}
\centering
\includegraphics[width=1.0\textwidth]{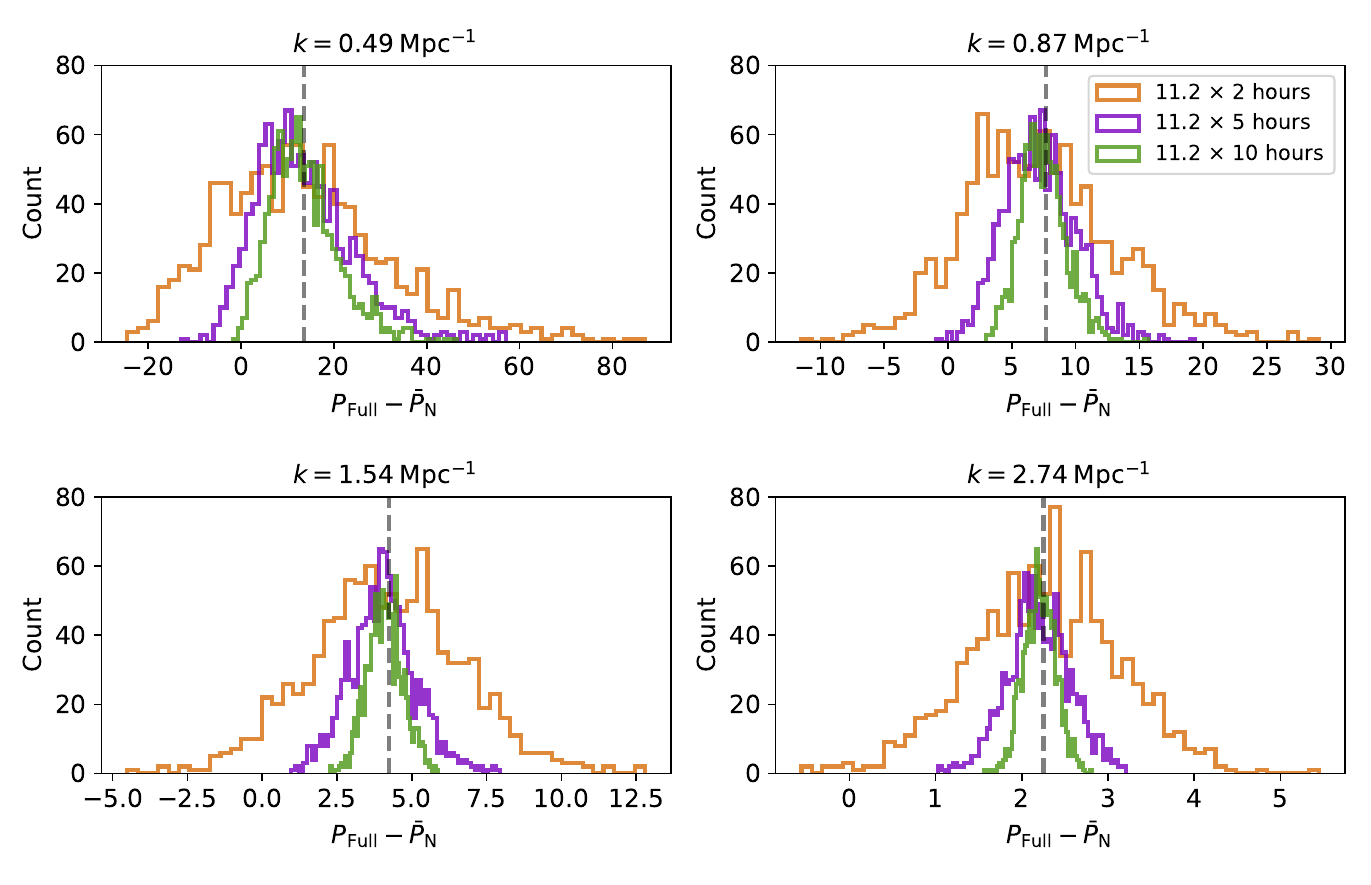}
\caption{Distribution of 1d power from $1000$ realizations for the first four $k$ values after subtracting the average noise spectrum as histograms. The expected values of the 21cm power spectrum are shown as vertical dashed lines in each subplot. The power on the x axis has units of $\rm{mK}^2 \rm{Mpc}^3$).}
\label{hist}
\end{figure*}

\begin{figure*}
\centering
\includegraphics[width=0.9\textwidth]{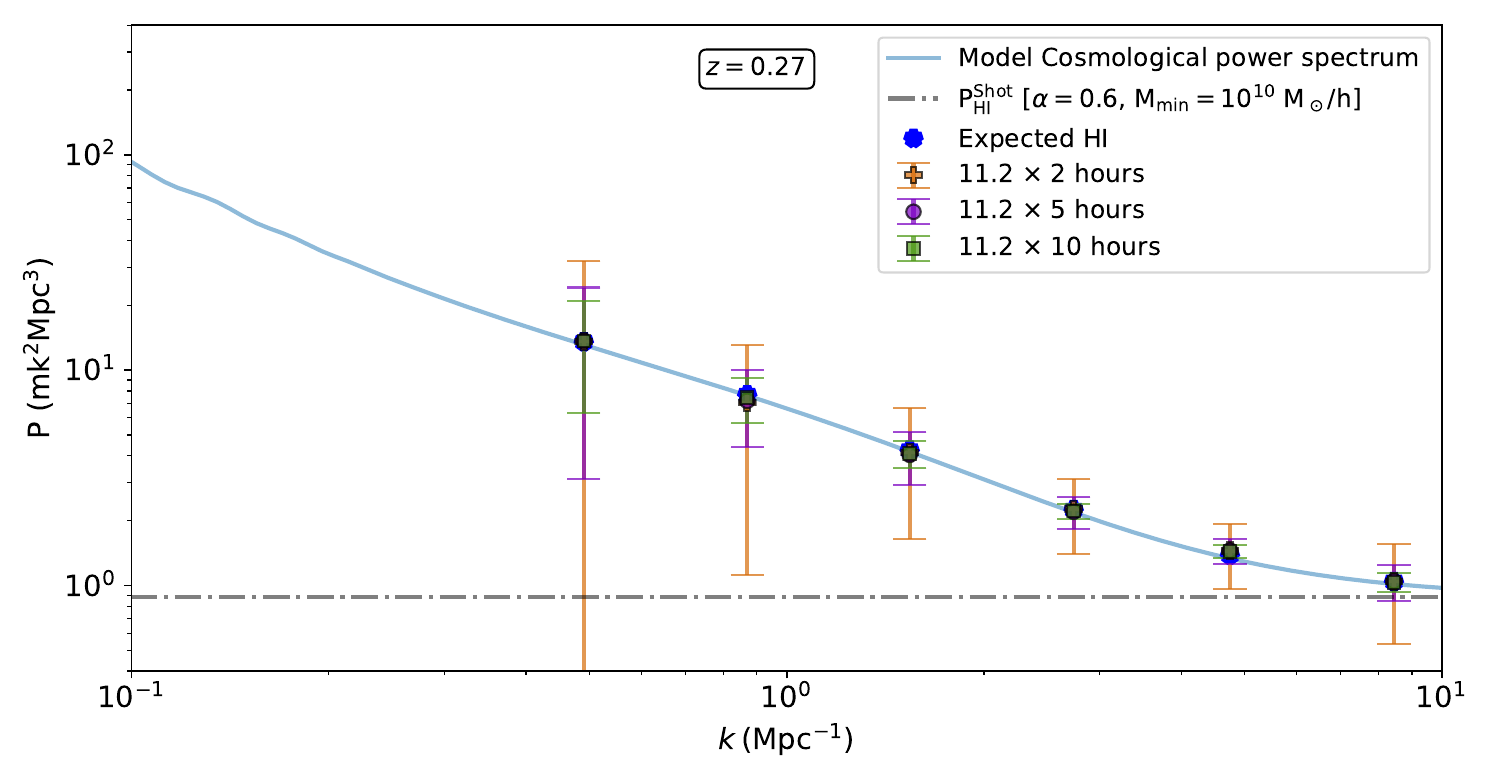}
\caption{Extracted HI power spectrum with $1\sigma$ errorbars from the distribution of average noise subtracted 1d power spectra (\autoref{hist}). For comparison, the model cosmological signal with shot noise (appendix~\ref{shot_noise}) are included too.}
\label{1d_result}
\end{figure*}

 To assess the prospect of detection in detail, we investigate scenarios where we increase the duration of observation in integer multiples of the existing $11.2$ hours case. We implement the increased integration hours by assuming observation on the same field taken at the same time as the existing data on different days, i.e. we do not gain any additional \textit{uv} points. Then the data can be coherently added on the same \textit{uv} point from multiple days which increases the sensitivity at that \textit{uv} point without any loss of signal. We consider three hypothetical cases where we consider $2$, $5$ and $10$ times of the existing $11.2$ hours of tracking observation on the COSMOS field. For these three cases, we also generate 1000 realizations of power spectrum from the simulation pipeline. In these 1000 estimates of power spectrum, the input HI and noise components are independent realizations with the same foreground contribution. As the foreground wedge is excluded from calculating the 1d power, the foreground contribution is suppressed, and it includes dominant contributions from both the cosmological signal and system noise. 
 
 To extract the cosmological 21cm signal from the 1d power spectra, we can subtract a good thermal noise model from it. However, a single realization of thermal noise may not capture the thermal noise properties well and therefore, we generate the thermal noise power spectra from $1000$ realizations and obtain a thermal noise model by calculating the mean for all three cases. The mean thermal noise power can then be subtracted from the 1d realizations, and we can have estimates of the noise-free 21cm signal. In \autoref{hist}, we present the distribution of 1d power after subtracting the average noise power spectra from full simulations as histograms for the first four k values. If the noise model is accurate, \autoref{hist} should manifest the distribution of the extracted cosmological signal which exhibits a Gaussian profile for all the cases considered here. The histogram curve gets narrower with increasing integration time as the variance decreases due to decrement in thermal noise power. In \autoref{1d_result}, we present the mean of the distributions shown in \autoref{hist} with error bars which are calculated as the corresponding standard deviations. \autoref{1d_result} clearly indicates that we are able to extract the input cosmological HI signal with reasonable accuracy with our pipeline (no bias).

\begin{figure}
\centering
\includegraphics[width=0.5\textwidth]{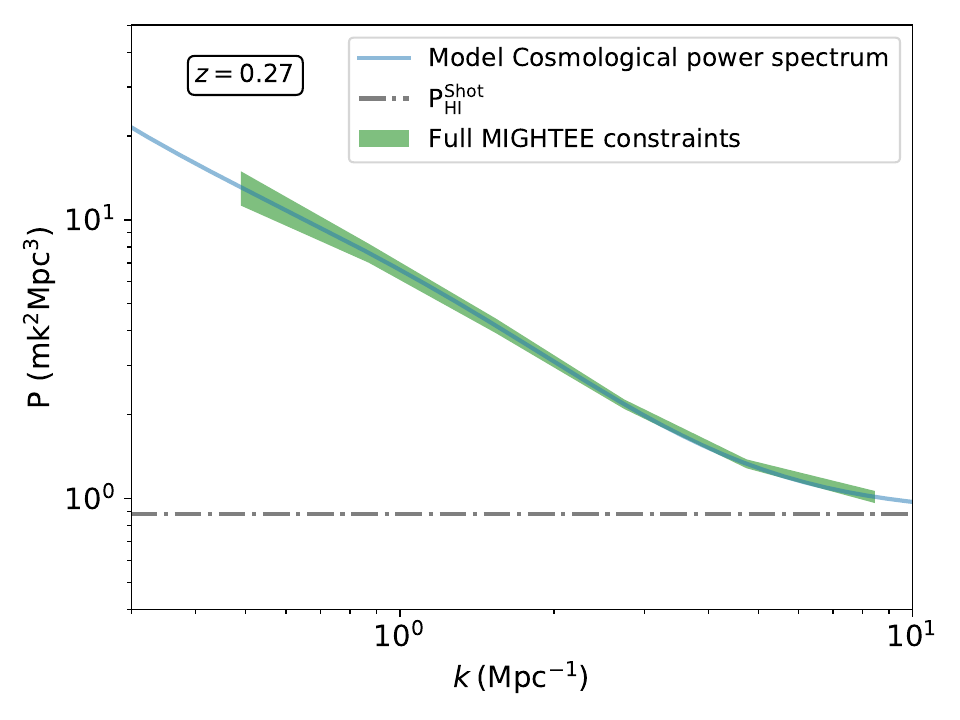}
\caption{Expected constraints with full MIGHTEE survey}
\label{full_mightee}
\end{figure}

So far, we have considered deep integration on a single pointing by increasing our fiducial observation case of $11.2$ hours. We have multiplied the integration time by $\rm{N_{mult}}=2,5,10$ and observed that the thermal noise power spectra drops as $P_{\rm{N}}(11.2 \:\rm{hours})/N_{\rm{mult}}$. However, in the absence of long observation on a single pointing, sensitivity can be improved by averaging power spectrum estimates from multiple independent fields. The MIGHTEE survey aims to study four well-known fields from the southern hemisphere: COSMOS, XMM-LSS, ECDFS and ELAIS-S1 \citep{Jarvis_2016} with multiple pointings to cover $20$ degree$^2$ sky area. When the power spectrum measurements from independent fields are combined, the uncertainty in the mean power spectrum reduces as $\sqrt{\rm{N_{field}}}$; where ${\rm{N_{field}}}$ is the number of independent fields. The MIGHTEE survey aims to achieve $\sim 2 \mu$Jy sensitivity over the L-band. This can be achieved with approximately 1000 hours of integration distributed across all four fields. In \autoref{pow_est_table}, we provide the power spectrum constraints that can be achieved with single COSMOS pointing of $22.4$ hours, as well as the full MIGHTEE survey (\autoref{full_mightee}). For the full survey, we have assumed uniform distribution of independent pointings across the full survey area. Please note that the actual sensitivity will depend on the observation strategy such as number of independent pointing and integration time per pointing. In \autoref{pow_est_table} and \autoref{full_mightee}, we provide ballpark estimates of the full MIGHTEE survey capability. These results give a total signal-to-noise ratio (SNR) $\approx 4.92$ on the COSMOS field (22.4 hours integration) in the range $0.3<k<11$ Mpc$^{-1}$; whereas the full MIGHTEE survey is capable to achieve a SNR $\approx 49$ for the same $k$ range. These estimates are calculated for the particular HI model used in this paper (\autoref{HI_model}), and a $20\%$ variation in $\bar{T}_b(z) b_{\rm HI}(z)$ can cause the SNR to vary approximately $20 - 40\%$ from the estimated values across the $k$ range explored in this paper depending on the contribution from noise.

\begin{table}
\centering
\begin{tabular}{| c | c | c c |} 
 \hline
  & & \multicolumn{2}{c |}{Power spectrum ($1\sigma$ estimates)} \\
 $k$ & $P_{\rm{HI}}(k)$ & \multicolumn{2}{c}{(mK$^2$Mpc$^3$)} \\
 (Mpc$^{-1}$) & (mK$^2$Mpc$^3$) & 22.4 hours & Full MIGHTEE \\
  & & (COSMOS) & (1000 hours, 4 fields)\\
 \hline
 0.49 & 13.55 & 18.428 & 1.846 \\
 0.87 & 7.68 & 5.945 & 0.596 \\
 1.54 & 4.25 & 2.508 & 0.251 \\
 2.74 & 2.25 & 0.856 & 0.086 \\
 4.75 & 1.38 & 0.489 & 0.049 \\
 8.45 & 1.04 & 0.507 & 0.051 \\
 \hline
\end{tabular}
\caption{Table summarizing the estimated power spectrum constraints with the MIGHTEE survey.}
\label{pow_est_table}
\end{table}

\section{Conclusions and Future Scope}
\label{section_conclusion}
Intensity mapping of the neutral Hydrogen line is a promising avenue to probe the large scale structure of the Universe and provide precision cosmological constraints. MeerKAT with its single dish capabilities can probe scales $> 1$ degree. As an interferometric array, MeerKAT has a dense core and therefore can also probe the quasi-linear cosmological scales, complementary to the single-dish intensity mapping. MIGHTEE, one of MeerKAT's large survey projects, aims to provide radio continuum, spectral line and polarisation information. In this paper, we have shown that MeerKAT has the unique capability to make a statistical detection of HI on quasi-linear scales, the first of its kind, with the existing and planned MIGHTEE observations on well-known fields.

We have developed a simulation pipeline that can emulate the MIGHTEE observations. The current simulation includes contributions from the HI signal, noise and extragalactic point sources on the COSMOS field. With the current setup, we show that the full MIGHTEE survey will have high sensitivity to detect the HI power spectrum at $z = 0.27$ for $k > 0.3$ $\rm{Mpc}^{-1}$. We have shown the constraints that can be achieved on the HI power spectrum with a single pointing deep integration on the COSMOS field. As MIGHTEE includes observations from multiple fields, the power spectrum can also be estimated independently from those fields and combined to further reduce the uncertainty.

There are ongoing efforts with dedicated pilot surveys in MeerKAT single-dish mode with the goal of first detection of HI power spectrum on large scales. In this mode, we are constrained to scales $>1$ degree and it will allow us to set important constraints on the Baryon acoustic oscillations and resdshift space distortion. Given that MeerKAT baselines are not small enough to probe large cosmological scales, we are constrained to quasi-linear scales in the interferometric mode. The high signal-to-noise measurements with MIGHTEE on these scales will be useful to determine the HI content of galaxies, the correlation between HI content and star formation rates \citep{Wolz_2016}; and break the degeneracy
between $\Omega_{\rm HI}$ and $b_{\rm HI}$ \citep{Chen_2021}. A successful detection of HI on quasi-linear scales will also allow to better understand various cosmological quantities such as the spectral index of fluctuation, constraints on warm dark matter, the quasi-linear redshift space distortions and the measurement of the HI content of the Universe up to $z\sim 0.5$ along with the direct comparison with HI simulations such as SIMBA \citep{Dave_2019}.

The next step will be to attempt to detect the signal with the available and upcoming data across a range of redshifts from all four fields identified for the MIGHTEE project. As we go deeper in sensitivity, further improvements might be necessary to the simulation in order to validate the signal extraction (e.g. the effects of the MeerKAT primary beam on the HI signal contamination). We therefore plan to continuously improve this simulation pipeline using insights from the calibrated data.

This paper is focused on developing the simulation pipeline and testing how well the MeerKAT interferometer would be able to constrain the HI power spectrum. It is an important step since MeerKAT wasn’t developed with this in mind (as compared to highly redundant interferometers). This work demonstrates that MeerKAT is capable of the first ever detection of the HI IM power spectrum in auto-correlation with $< 100$ hours of observation with the current configuration. Given some of the leakage issues, further improvements on the calibration pipeline will be required. Simply excluding the scales not consistent with thermal noise is not enough with the current data. We are in the process of analysing a larger dataset now which, combined with some improvements in calibration and flagging, will hopefully lead to a detection of the power spectrum.

\section*{Acknowledgement}
We thank the anonymous referee for the helpful comments
on the draft that have contributed to improving this paper.
MGS and SP acknowledge support from the South African Square Kilometre Array Project 
and National Research Foundation (Grant No. 84156).
We acknowledge the use of the Ilifu cloud computing facility - www.ilifu.ac.za, through the Inter-University Institute for Data Intensive Astronomy (IDIA). This work was carried out using the data processing pipelines developed at the IDIA and available at https://idia-pipelines.github.io. NM acknowledges support from the Bundesministerium f{\"u}r Bildung und Forschung (BMBF) award 05A20WM4.
The MeerKAT telescope is operated by the South African Radio Astronomy Observatory, 
which is a facility of the National Research Foundation, an agency of the Department 
of Science and Innovation.

\section*{Data availability}
The data underlying this article will be shared on reasonable request to the corresponding authors (SP, MGS).

\bibliography{references}
\appendix
\section{Shot Noise}
\label{shot_noise}
The shot noise power due to Poisson fluctuation in halo number is given by \citep{Bull_2015}:
\begin{equation}
P_{\rm HI}^{\rm shot}(z) = \left(\frac{\bar{T}_b (z)}{\rho_{\rm HI}(z)}\right)^2 \int_{M_{\rm min}}^{M_{\rm max}}dM \frac{dn}{dM} M_{\rm HI}^2(M).
\label{shot_noise}
\end{equation}
The HI mass is modelled as $M_{\rm HI}(M) = AM^\alpha$, with $\alpha=0.6$ and $A\sim 220$ \citep{Santos_2015, Santos_2017} consistent with the HI model used for the signal component in the power spectrum simulation.
\end{document}